# INFLUENCE OF HUMIDITY ON MICROTRIBOLOGY OF VERTICALLY ALIGNED CARBON NANOTUBE FILM

V. TURQ [1], N. OHMAE [2], J. M. MARTIN [1], J. FONTAINE [1], H. KINOSHITA [2], J. L. LOUBET [1]

[1] *Ecole Centrale de Lyon, Laboratoire de Tribologie et Dynamique des Systèmes, UMR CNRS 5513, 69134 Ecully Cedex, FRANCE*
[2] *Department of Mechanical Engineering, Faculty of Engineering, Kobe University, Rokkodai, Nada, Kobe 657-8501, JAPAN*

*Abstract:*

The aim of this study is to probe the influence of water vapor environment on the microtribological properties of a forestlike vertically aligned carbon nanotube (VACNT) film, deposited on a silicon (001) substrate by chemical vapor deposition. Tribological experiments were performed using a gold tip under relative humidity varying from 0 to 100%. Very low adhesion forces and high friction coefficients of 0.6 to 1.3 resulted. The adhesion and friction forces were independent of humidity, due probably to the high hydrophobicity of VACNT. These tribological characteristics were compared to those of a diamond like carbon (DLC) sample.

*Keywords:* Carbon nanotubes, nano- and microtribology, diamond like carbon, hydrophobicity

## 1  Introduction

Carbon nanotubes (CNTs) can be defined as a network of $sp^2$ hybridized carbon atoms, consisting of single or multilayers, whose length can reach a few micrometers. Carbon nanotubes were shown to exist in 1991 by S. Iijima [1]. Since then a lot of research has been conducted to characterize them and to apply their outstanding mechanical and electronic properties in industrial applications. Experimental and simulation results demonstrate that CNTs have a high Young's modulus and stiffness in the direction of the nanotube axis, as well as considerable mechanical strength, and great flexibility perpendicular to the axis. Therefore CNTs have important technological potential in materials science, in microscopy, in microelectronics and as micromachine elements. Nanobearings [2], CNT- based composites using polymer-, ceramic- or metal-matrices [3], or a diamond thin film-matrix [4] have been successfully produced. Nevertheless, the tribological properties of carbon nanotubes have seldom been studied on the nano- or microscale.

Classical molecular dynamics simulations were used by Ni and Sinnott [5] to predict the tribological behavior of a nanotube bundle settled between two (111) hydrogenated diamond surfaces, for various patterns of bundles. Vertical assembly of six closed nanotubes, about 25Å long, was one of the computed bundles. At a contact pressure of 1.44 GPa on the surface, stick-slip phenomena were observed. At higher pressure (11.5 GPa), the nanotubes bent under the effect of the shear and then returned to the initial shape. The friction coefficient calculated for the weaker pressure is 2.1 while that calculated for the higher pressure is 0.87.

First experiments on CNT were reported by Ohmae et al. [6]. They used one gold tip (radius of curvature around 3 µm) and also one silicon tip perpendicularly covered by nanotubes, to probe in ambient air a forestlike vertically aligned nanotube film deposited on a Si (001) surface. The relative humidity was not controlled. The friction coefficient was 0.5 for the nanotube / gold tip experiments. However, a friction coefficient lower than 0.1 was achieved for the nanotube / nanotube experiment.

The aim of this paper is to analyze the tribological behavior of vertically aligned carbon nanotube (VACNT) films further, and to better understand the effect of the environment on their tribological properties. Therefore experiments were conducted with different forces (µN range), using a gold tip (radius 20 µm), under controlled relative humidity varying from 0 to 100%, at room temperature. The results were compared to those of a hydrogenated diamond like carbon (DLC) sample.

## 2  Experiments

VACNT were synthesized on a Si (001) wafer by microwave plasma enhanced chemical vapor deposition (CVD). Prior to the CVD process, a thin iron catalyst film was deposited on the Si substrate. Annealing was then conducted to enable the formation of Fe catalytic droplets on the surface. The reaction gases used for the synthesis were $CH_4$ and $H_2$, with flow rates of 100 and 10 sccm respectively and at a total pressure of 2.7 kPa. The microwave plasma was ignited at 2.45 GHz and 500 W. The films were obtained after 40 seconds under a bias voltage of –500 V and a temperature of 700°C. The synthesized VACNT film consisted of 6 µm long carbon nanotubes (Figure 1).

Transmission electron microscopy (TEM) observations showed that multiwalled carbon nanotubes grew on the silicon wafer with a diameter of approximately 20 nm. The roughness of the VACNT surface was relatively high, the root-mean-square height equaling 105 nm on the micrometer scale (tapping-mode atomic force microscopy measurement). Note that the gold tip could easily be in contact with several nanotubes because of the large radius of the tip (about 20 μm) compared to the nanotube diameter.

The laboratory-made microtribometer used a rectangular cantilever and an optical lever system to measure normal and tangential forces, as a conventional Atomic Force Microscope (AFM) does (Figure 2-a). The probe tip was glued to a laboratory-made aluminum cantilever. The probe used was a gold tip with an apex radius of 20 μm, prepared by electropolishing with KOH electrolyte. The forces acting on the cantilever were calibrated as is done for classical tribometers [7], aluminum wire weights being hung on the calibration hook (Figure 2-b). The weight versus horizontal and vertical deflection curves enabled the normal and lateral spring constants of the cantilever to be determined. The microtribometer was installed in a vacuum chamber emptied by rotary and turbo-molecular pumps. The lowest vacuum in the chamber was $10^{-5}$ Pa. A capacitance-type humidity meter was used to measure relative humidity. Force-displacement curves and friction measurements were obtained at several humidity levels and different surface locations. The adhesion force ($F_0$) was determined at room temperature from force-displacement curves, $F_0$ being equal to the maximum pull-off force. The real normal load (N) is considered to be the sum of the applied load (F) and the adhesion force ($N = F + F_0$) [8].

Friction experiments were conducted at room temperature with an amplitude of 10 μm and a sliding speed of about 8 μm/s. Different loads (N=0 to ~30 μN) were used at each given humidity level, with at least 4 cycles of reciprocating motion. Two cases were considered: either the cantilever horizontal deflection (X) had a periodic triangular shape, or X had a periodic rectangular shape. In the first case, sticking occurred between probe and surface. There was no sliding and friction coefficients could not be determined. In the second case, sliding did occur. The mean amplitude of the periodic rectangular signals could be linked to the friction force T. The apparent mean friction coefficients, $\mu_{app}$, defined by the ratio between T and N ($\mu_{app} = T/N$) were first calculated for all normal loads. The real friction coefficient $\mu$ and the shear yield force $T_0$ were determined using the following relation: $T - T_0 = \mu \times N$.

## 3 Results

Figure 3-a shows the adhesion force $F_0$ between the VACNT film and the gold tip as a function of relative humidity. In this case, $F_0$ was hardly measurable on the force-displacement curve. Figure 3-b shows the adhesion force $F_0$ between DLC coating and the gold tip as a function of relative humidity. Unlike the VACNT film, the adhesion force for DLC depends on the humidity level. Very small adhesion forces were observed at low humidity levels and high adhesion at high relative humidity. The transition occurs around a relative humidity of 60%.

The microtribological results obtained with the gold tip and the VACNT film indicated very high apparent ($\mu_{app}$) as well as real ($\mu$) friction coefficients, lying between 0.58 and 1.31. The mean value for $\mu_{app}$ is 0.98 (Figure 4-a and Figure 5-a). Within the experimental scatter, the effect of the water vapor environment on friction coefficient cannot be easily recognized. But the shear yield force ($T_0$) could also be described as a function of the relative humidity (Figure 5-a). Unlike the friction coefficients, $T_0$ seems to depend on the humidity level. It varies between -0.21 and 0.96 µN. Negative values have no physical meaning but are due to the scattering of results. Furthermore, it can be noticed that the scattering of the results is specific to carbon nanotubes. For instance, the apparent friction coefficient maximum deviation is around 0.44 for the gold / VACNT contact while it is around 0.22 for the gold / DLC contact.

Figure 4-b depicts the apparent friction coefficient as a function of the relative humidity in the case of the gold / DLC contact. It stays at a relatively stable level around 0.18. The real friction coefficient lies between 0.05 and 0.18, as shown in Figure 5-b. The yield shear force $T_0$ is about 0.5 µN.

## 4 Discussion

VACNT film does not exhibit the same type of adhesion behavior as the hydrogenated DLC surface tested under the same conditions and as the Ni, Cu, and Au samples studied in similar Atomic Force Microscope (AFM) experiments [7]. At low humidity levels for DLC, as well as for Ni, Cu, and Au samples, adhesion forces are very small. However at around 60% relative humidity, the adhesive forces increase sharply. The main reason for this behavior is probably linked to the formation of capillary bridges or water meniscus between the tip and sample surfaces, resulting in a reduced Laplace pressure. When the surfaces are hydrophilic, it generates an additional normal force, which increases the adhesion between tip and sample. The remarkably weak adhesion of the VACNT film at any relative humidity could therefore be attributed to the hydrophobicity of nanotubes. Figure 6 represents the contact angle measurement of a water droplet on a VACNT film. The contact angle measured was about 160 degrees. The rough surface of the VACNT film combined with the hydrophobic nature of a graphite surface may account for such large contact angles. Thus, dewetting phenomena can occur and no adhesive force due to the water meniscus formation can be recorded.

The contact adhesion between a tip and a carbon-based material surface is generally well described [9,10] by the Derjaguin Muller and Toporov (DMT) model [11]. In this model the adhesion force $F_0$ is calculated by the following expression:

$$F_0 = 2\pi R \Delta \gamma$$

Where $\Delta\gamma$ denotes the adhesion energy between the tip and the solid material and R the mutual radius of curvature, which is taken, in this case, to be the radius of the tip, i.e. 20 μm, taking the material surface to be flat. $\Delta\gamma = \gamma_1+\gamma_2-\gamma_{12}$, $\gamma_1$ and $\gamma_2$ are surface energies of the bodies 1 and 2 and $\gamma_{12}$ is the interfacial energy.

From this model, the adhesion energy corresponding to the experimental adhesion force in vacuum can be deduced. For DLC coating, in which $F_0$ is equal to 0.85 μN in vacuum, the expression leads to an adhesion energy of 6.8 mJ/m$^2$. The surface energy of DLC coatings is commonly reported in the 20-40 mJ/m$^2$ range for various DLC coatings ([12]), leading generally to higher adhesion energy. However, a 6 mJ/m$^2$ value for adhesion energy has been reported ([13]) between a lower surface energy DLC coating, produced by electron-beam graphite elaboration, and a commercial AFM NanoScope II $Si_3N_4$ cantilever. For the VACNT film, the experimental adhesion force value of 0.29 μN in vacuum leads to an adhesion energy of 2.3 mJ/m$^2$. This could account for the high hydrophobicity of the VACNT film.

Both apparent and real friction coefficients measured in the VACNT case were much higher than in the DLC case. SEM observations of the VACNT film and of the gold tip failed to provide evidence of a transfer film or wear track on the surface or damage to the tip. Studying transfer film formation and its relation to friction evolution is a key issue to understanding tribological behavior, and is a possible future development of this work.

Direct comparison between experimental friction coefficients obtained in this study and those predicted by Ni and Sinnott [5] is difficult. Indeed, the pressures applied in their simulations are very high (1.44 and 11.5 GPa), compared to those of our experiments (less than 0.5 GPa) and the length of carbon nanotubes is also very much smaller. However, as in the simulations, high friction values were found. A possible explanation for these very high friction coefficients is the existence of high repulsive forces applied to the tip by the nanotubes during horizontal displacement, and due to the bending of the carbon nanotubes [14,15].

When friction dependence on contact pressure is found [16], the yield shear force evaluation approach is especially pertinent. Indeed, the yield shear force $T_0$ is not equal to 0 in our experiments, either for VACNT, or for DLC. Two classical hypotheses can explain this behavior. The first one is the existence of adhesion forces between tip and sample. The second is linked to cohesion forces inside the contacting interface [17]. In our experiments, the origin of the normal force (N) was taken at the minimum of the force in the force-displacement curve. Thus, in this study, the contribution of the adhesion force to the yield shear force $T_0$ could only be attributed to uncertainty in the measurements. When adhesion forces were taken into account, the results were applicable to the second case in which the mechanical cohesion of the interface is considered. For DLC, transfer film formation is a well known phenomenon, so the yield shear force could be a characteristic of the transfer film. For VACNT film, since no transfer layer could be detected, cohesive forces inside the film, i.e. between nanotubes, have to be considered. For instance, van der Waals forces, or π-π* orbital overlap between carbon nanotube walls could account for such cohesive forces. A rough estimation of yield shear force $T_0$ based on the non-retarded van der Waals force between two parallel cylinders [18] could be conducted. The non-retarded van der Waals interaction free energy W between two parallel cylinders with same radius R is given by the following expression:

$$W = \frac{AL}{24D^{3/2}}\sqrt{R}$$

Where A denotes the Hamaker constant of the material, D the distance between the two cylinders and L the length of the cylinders.

This leads to the following f interaction force:

$$f = \frac{AL}{16D^{5/2}}\sqrt{R}$$

For the VACNT coating, taking the Hamaker constant for the CNT wall to be the one of graphite, i.e. $23.8\times10^{-20}$ J, the length of the CNTs to bo 6 μm, the radius of a CNT to be 10 nm and the distance between two nanotubes to be 10 nm, the expression leads to an interaction force of $8.93\times10^{-10}$ N. Thus, a yield shear force $T_0$ around 0.54 μN, as the one obtained at 0% relative humidity for the VACNT film, could be generated by van der Waals forces between about 600 nanotubes pairs. This would give the number of nanotubes involved in the contact along our friction experiments. More sophisticated estimates of the micro-mechanism accountable for the global friction behavior should be made, as well as more precise experiments, to confirm this hypothesis. For instance, the effect of the length of the nanotubes, of their diameter and of their surface coverage should be considered.

## 5 Conclusion

The vertically aligned carbon nanotube (VACNT) film exhibited very low adhesion and surprisingly high friction, at any relative humidity. This can be attributed to the high hydrophobicity of the carbon nanotubes. For less hydrophobic DLC coatings the adhesion forces vary with humidity. However friction seems to be independent of relative humidity.

Furthermore, the strong cohesion forces existing in the film between the nanotubes may relate to this anomalous tribological property. A more detailed study of the influence of the thickness and density of the VACNT film on the tribological behavior will be required to clarify this issue.

# 6 Acknowledgements

This work was funded by the *Direction de l'Enseignement Supérieur de la Région Rhône-Alpes* thanks to a *Bourse régionale de formation à l'étranger*, the French *Ministère de l'Education Nationale, de l'Enseignement Supérieur et de la Recherche*. We would also like to thank Ippei Kume and Roselina Nik from the *Laboratory for Surface, Interface and Tribology of Kobe University* for their help, for AFM tapping mode analyses and SEM pictures and Prof. Denis Mazuyer from the *LTDS of Ecole Centrale de Lyon* for helpful discussions.

# 7 References

[1] S. Iijima, Nature 354 (1991), 56.
[2] J. Cumings and A. Zettl, Science 289 (2000), 602.
[3] E. T. Thostenson, Z. Ren, T.-W. Chou, Composites Science and Technology 61 (2001), 1899.
[4] H. Schittenhelm, D. B. Geohegan, G. E. Jellison, Applied Physics Letters 81 (2002), 2097.
[5] B. Ni, S. B. Sinnott, Surface Science 487 (2001), 87.
[6] J. M. Martin, Presentation AVS 49th International Symposium (2002), Ultralow Friction Coatings and Surfaces.
[7] Y. Baba, H. Kinoshita, M. Tagawa, M. Umeno, and N. Ohmae, Japanese Journal of Tribology, vol. 44-6 (1999), 255.
[8] B.V. Derjaguin, V.V. Karassev, N.N. Zakhavaeva and V.P. Lazarev, Wear vol. 1 (1958), 277.
[9] M.J. Adams, B.J. Briscoe, J.Y.C. Law, P.F. Luckham and D.R. Williams, Langmuir 17 (2001), 6953.
[10] B.J. Briscoe, J. Law, D.R. Williams and P.F. Luckham, Int. Journal of Adhesion and Adhesives, vol. 14, n°2 (1994), 77.
[11] B.V. Derjaguin, V.M. Muller and Y.P. Toporov, J. Colloid Interface Science 53 (1975), 314.
[12] M. Grischke, A. Hieke, F. Morgenweck, H. Dimigen, Diamond and Related Materials 7 (1998) 454.
[13] V. Snitka, A. Ulcinas, M. Rackaitis, D. Zukauskas, M. Fukui, Diamond and Related Materials 6 (1997), 1.
[14] H.J. Qi, K.B.K. Teo, K.K.S. Lau, M.C. Boyce, W.I. Milne, J. Robertson, K.K. Gleason, J. Mechanics and Physics of Solids 51 (2003), 2213.
[15] H. Kinoshita, I. Kume, M. Tagawa and N. Ohmae, Applied Physics Letters, vol. 85, n°14 (2004), 2780.
[16] B.J. Briscoe, A.C. Smith, ASLE Transactions, vol. 25, 3 (1982), 349.
[17] D. Mazuyer, J. M. Georges and B. Cambou, J. Phys. France 49 (1989), 1057.
[18] J. Israelachvili, Intermolecular & Surface Forces (Academic Press Limited, London, 1992).

Figure captions

Figure 1: SEM photograph of vertically aligned carbon nanotubes (VACNTs) grown on thermally oxidized silicon (001) substrate.

Figure 2: (a) AFM layout used in this study; (b) schematic diagram of the cantilever and tip unit.

Figure 3: (a) Adhesion force versus humidity curve for VACNT film; (b) adhesion force versus humidity curve for DLC coating at three different times. The solid line, representing the mean adhesion force, is a guideline for the eyes.

Figure 4: (a) Apparent friction coefficient $\mu_{app}$ versus relative humidity for VACNT film, the solid line represents the linear fit of the curve: 0.98 is the value of the intercept point and $4\times10^{-4}$ is the slope; (b) apparent friction coefficient $\mu_{app}$ versus relative humidity for DLC coating, the solid line also represents the linear fit: 0.18 is the value for the intercept point and $5\times10^{-5}$ is the slope.

Figure 5: (a) Real friction coefficient $\mu$ and yield shear force $T_0$ versus relative humidity for VACNT film; (b) real friction coefficient $\mu$ and yield shear force $T_0$ versus relative humidity for DLC coating.

Figure 6: Optical photograph showing water droplet on VACNT film; the measured contact angle is about 161°.

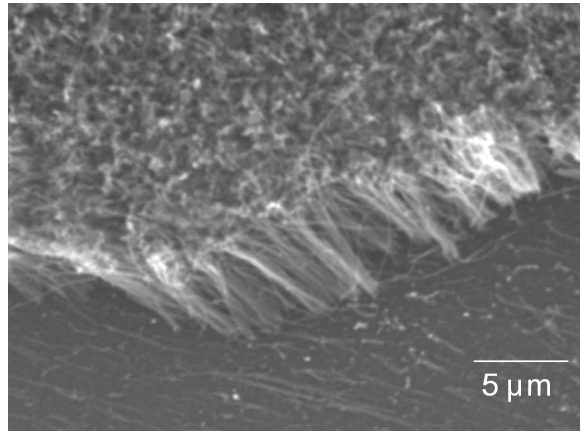

**Figure 1: SEM photograph of vertically aligned carbon nanotubes (VACNTs) grown on thermally oxidized silicon (001) substrate.**

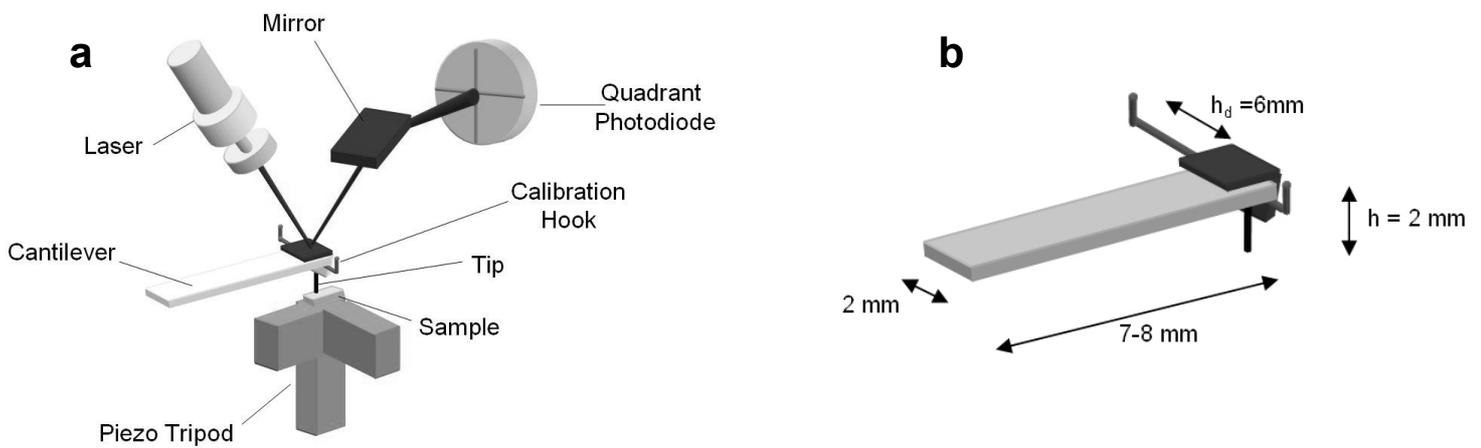

**Figure 2: (a) AFM layout used in this study; (b) schematic diagram of the cantilever and tip unit.**

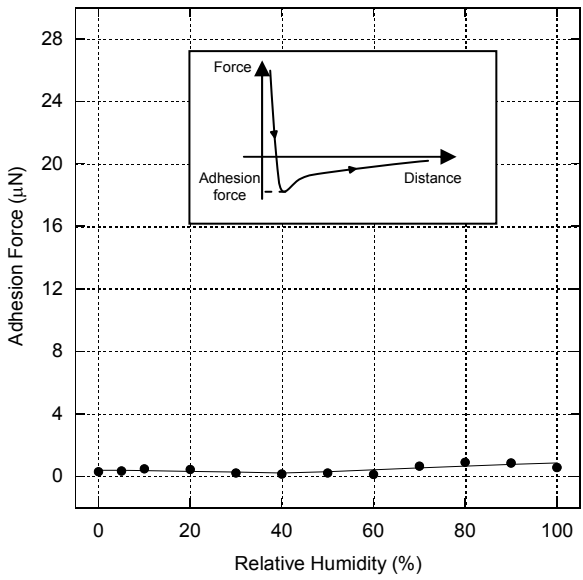
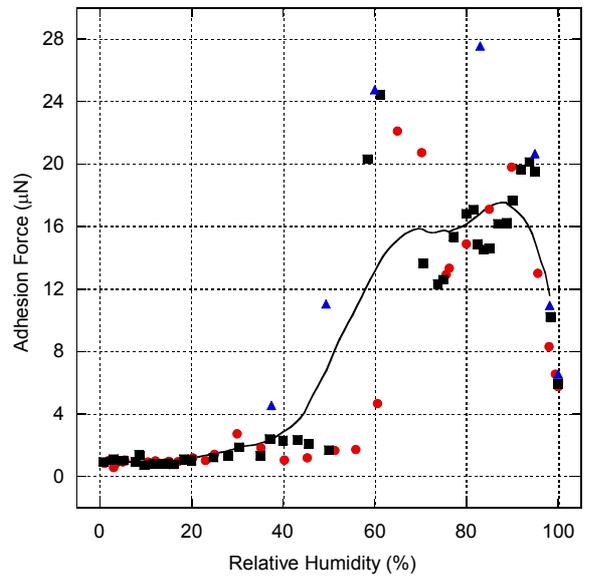

**Figure 3: (a) Adhesion force versus humidity curve for VACNT film; (b) adhesion force versus humidity curve for DLC coating at three different times. The solid line, representing the mean adhesion force, is a guideline for the eyes.**

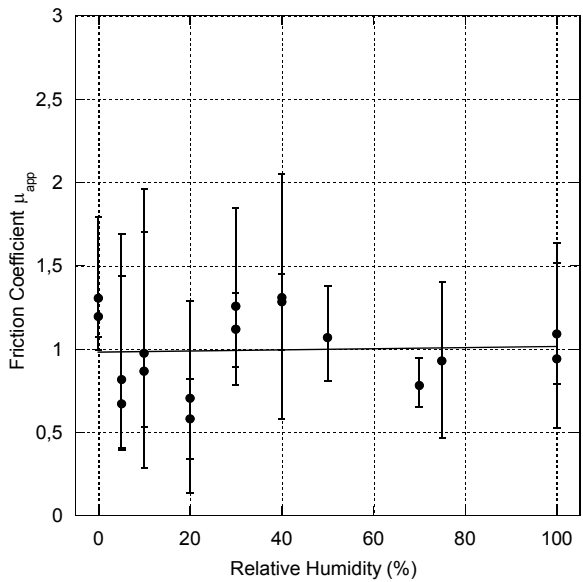
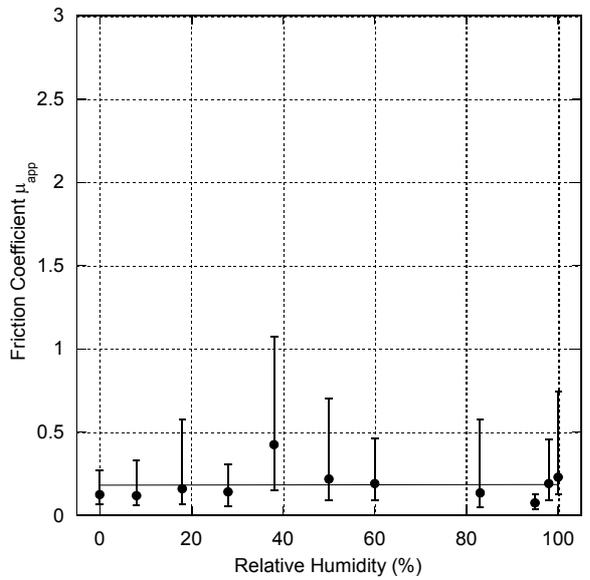

**Figure 4: (a) Apparent friction coefficient $\mu_{app}$ versus relative humidity for VACNT film, the solid line represents the linear fit of the curve: 0.98 is the value of the intercept point and $4\times10^{-4}$ is the slope; (b) apparent friction coefficient $\mu_{app}$ versus relative humidity for DLC coating, the solid line also represents the linear fit: 0.18 is the value for the intercept point and $5\times10^{-5}$ is the slope.**

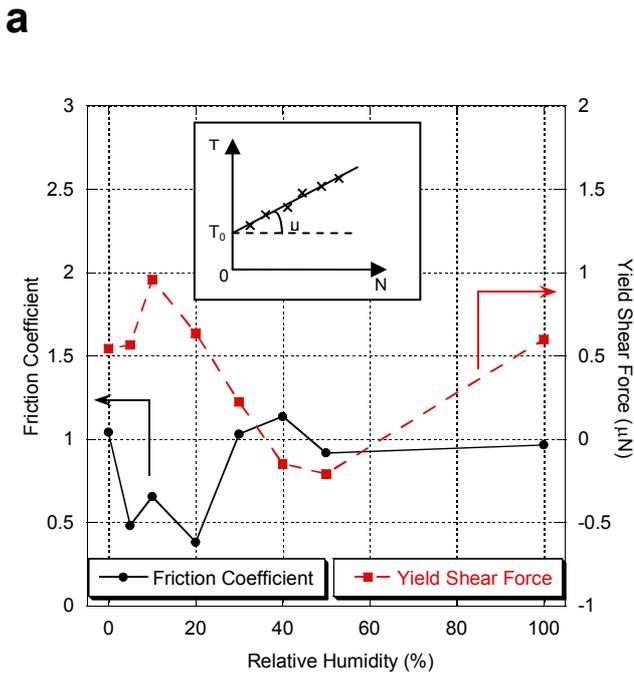 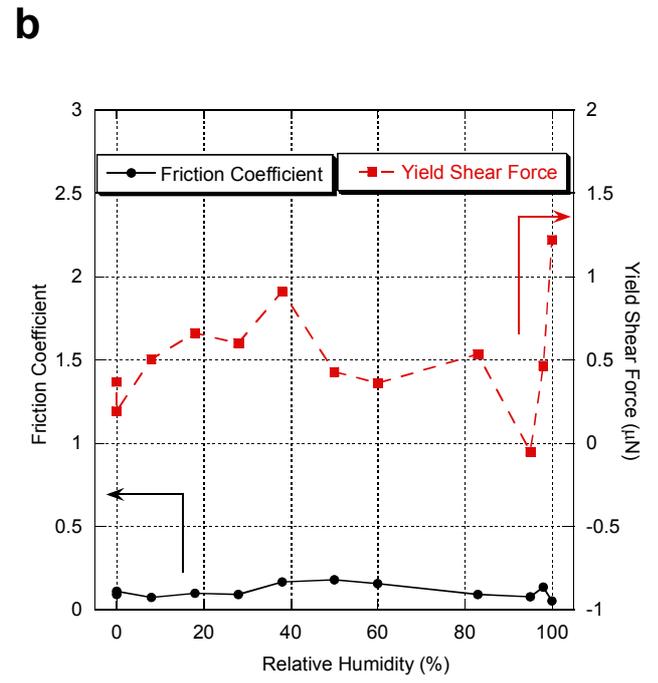

**Figure 5: (a) Real friction coefficient μ and yield shear force T$_0$ versus relative humidity for VACNT film; (b) real friction coefficient μ and yield shear force T$_0$ versus relative humidity for DLC coating.**

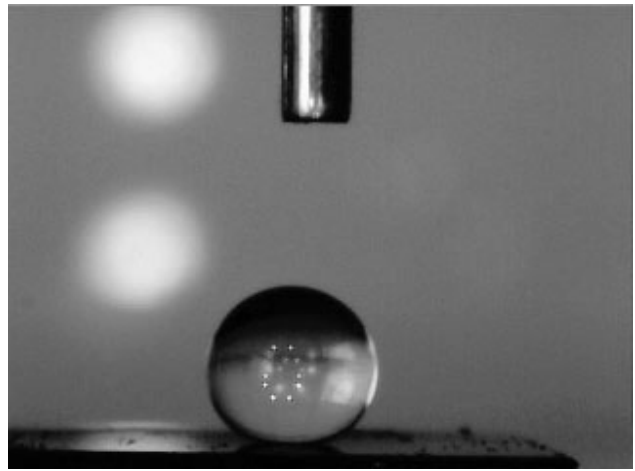

**Figure 6: Optical photograph showing water droplet on VACNT film; the measured contact angle is about 161°.**

12